\documentclass[twocolumn,showpacs,preprintnumbers,amsmath,amssymb]{revtex4}

\usepackage{graphicx}
\usepackage{dcolumn}
\usepackage{bm}
\usepackage{color}

\begin{document}



\title{Active Optical Clock
}

\author{Jingbiao Chen}\thanks{E-mail: jbchen@pku.edu.cn, phone: +86-10-6275-6853, Fax:
+86-10-6275-3208.}

\affiliation{Key Laboratory for Quantum Information and Measurements
of Ministry of Education, School of Electronics Engineering $\&$
Computer Science, Peking University, Beijing 100871, P. R. China\\}

\date{\today}

\begin{abstract}
This letter presents the principles and techniques of active optical
clock, a special laser combining the laser physics of one-atom
laser, bad-cavity gas laser, super-cavity stabilized laser and
optical atomic clock. As an example, a compact version of active
optical clock based on thermal Strontium atomic beam shows a
quantum-limited linewidth of 0.51 Hz, which is insensitive to laser
cavity-length noise, and may surpass the recorded narrowest 6.7 Hz
of Hg ion optical clock and 27Hz of very recent optical lattice
clock. The estimated 0.1Hz one-second instability and 0.27Hz
uncertainty are limited only by the relativistic Doppler effect may
be improved to 10mHz by using cold atoms.
\end{abstract}


\maketitle

In 1958, in a well-know paper\cite{Schawlow}, Schawlow and Townes
proposed to extend the maser techniques to laser. Just two years
later, the first laser was build by Maiman, also in this year,
Goldenberg, Kleppner, and Ramsey invented Hydrogen
maser\cite{Goldenberg}, an active microwave atomic clock, for which
scientists have enjoyed its excellent stability in a variety of
applications since its invention. However, we never have an active
optical clock so far. All the optical atomic clocks up to date, are
not working in active mode\cite{Gill,4,5}. Here, 45 years after the
invention of active microwave Hydrogen clock, this letter presents
the principles and techniques of active optical clock, which is the
optical frequency counterpart of active Hydrogen clock. It's a
special laser combining the laser physics of one-atom laser[6,7],
bad-cavity gas laser[8,9], super-cavity stabilized laser[10-12] and
passive optical atomic clock[3-5,13,14]. A compact version based on
thermal Strontium atomic beam shows a quantum-limited linewidth of
0.51 Hz, which will surpass the recorded narrowest 6.7 Hz of Hg ion
clock[14] and 27Hz of very recent optical lattice clock[5]. The most
interesting point is, the frequency of this active optical clock is
insensitive to cavity-length noise, which is currently the
limitation of available narrow-linewidth laser light sources. The
estimated 0.1Hz one-second instability and 0.25Hz uncertainty are
limited by the relativistic Doppler effect may be improved by using
cold atoms.  The active optical clock provides a new way to optical
atomic clock and precision laser spectroscopy, and it also opens a
door to long-time coherence physics, say hundred-second even
thousand-second coherence laser physics, the long-time counterpart
of Attosecond physics.

Using the definition of $a=\Gamma_{cavity}/\Gamma_{gain}$ [8], here
$\Gamma_{cavity}$ is the cavity loss rate, $\Gamma_{gain}$ the
frequency gain bandwidth of laser medium, when $a<<1$, a laser is
working in the good-cavity limit, and in the bad-cavity regime while
$a\geqslant1$ . Then let's cut a cake into four quadrants as shown
in Fig.1. Chronologically, the second quadrant was tasted by Gordon,
Zeiger, and Townes in 1954 by building the first ammonia maser, then
in 1960, Maiman tasted the forth one by building the famous first
laser, Rudy laser. In the same year, Goldenberg, Kleppner, and
Ramsey invented the best know Hydrogen maser, belongs to the second
quadrant. The third quadrant was reached with the one-atom maser[15]
in 1985 for cavity quantum electrodynamics. How about the first
quadrant? Following a semiconductor device, a He-Ne $3.39\mu m$ gas
laser[8] went into this bad-cavity regime of $a=1.4$ in 1994, like
the red cherry on the cake across the $a=1$ line in Fig.1. All the
conventional lasers are working in the good-cavity regime, the forth
quadrant. In this letter, the laser will be pushed deep down into
the bad-cavity regime $a>>1$, the ``new continent'' of laser at
first quadrant in the Fig.1.

\begin{figure}
  \includegraphics[width=7cm]{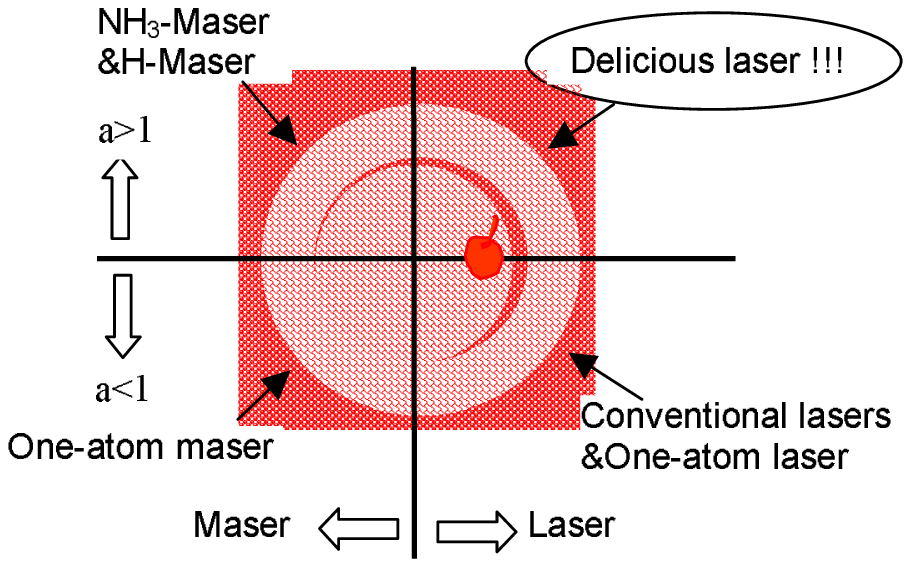}
  \caption{Birthday cake of Maser and Laser. Made by Townes, Basov,
and Prokhorov, and won them the Nobel Prize in physics1964. The
first quadrant is a ``new continent'' of lasers, good for active
optical clocks and laser spectroscopy with supper-narrow linewidth,
which is insensitive to cavity length variations.}\label{fig:1}
\end{figure}

For a homogeneously broadened single-mode laser, the quantum-limited
linewidth of a bad-cavity laser described by a modified
Schawlow-Townes formula[8],


\begin{eqnarray}
\Delta\nu_{laser}=\frac{\Gamma_{cavity}}{4\pi\overline{n}_{cavity}}N_{sp}{\left(\frac{1}{1+a}\right)}^2\nonumber\\
\left\{1+\left[\frac{4\pi\left(\nu-\nu_{0}\right)}{\Gamma_{gain}+\Gamma_{cavity}}\right]^{2}\right\},
\end{eqnarray}

 Here $N_{sp}=N_{p}/\left(N_{p}-N_{s}\right)$ is the spontaneous-emission factor
, $N_{s}$ , $N_{p}$ are the populations of the lower and upper
levels, $\nu-\nu_{0}$ is the detuning of the mode frequency $\nu$
from the center frequency $\nu_{0}$ of the gain profile, and
$\overline{n}_{cavity}$ is the steady-state number of photons in
laser mode. For an ideal four-level laser at zero detuning
($\nu=\nu_{0}$), Eq.(1) reduced to the Schawlow-Townes formula in
standard laser text books[16,17]
$\Delta\nu_{laser}=\Gamma_{cavity}/\left(4\pi\overline{n}_{cavity}\right)$
when laser operating in the good cavity limit ($a\ll1$  ). While
entering the bad-cavity regime, the results from HeNe 3.39$\mu$m gas
laser[8] agreed very well with theory as expressed Eq.(1). The
physics behind the factor $(1+a)^{-2}$ in Eq.(1) is the memory
effect of atomic polarization[8,9,18].

\begin{figure}
  \includegraphics[width=7cm]{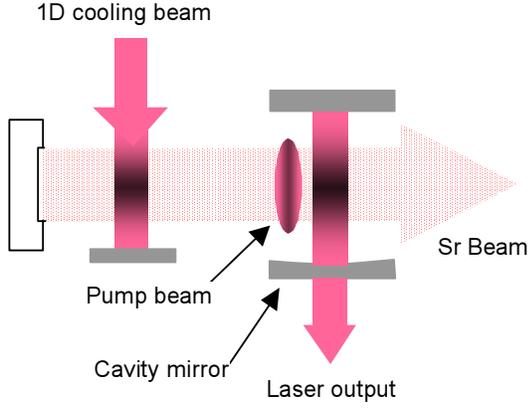}
  \caption{Strontium atomic beam laser. It is very similar with the
one-atom laser in structure, but with higher atomic beam flux,
larger cavity and lower atom-cavity coupling constant.}\label{fig:2}
\end{figure}

Let's construct a gas laser with thermal Strontium ($^{88}$Sr)
atomic beam by adjusting the one-atom $^{138}$Ba laser[6,7]
technically. The lasing transition is at the 689nm line of $5s5p$ $
^{3}P_{1}(m=0) - (5s)^{2}$ $^{1}S_{0}$, in which the $^{3}P_{1}$
state(lifetime $\tau_{sp}=21\mu$s ) has 7.6kHz decay rate to the
$^{1}S_{0}$ state [19]. The first adjustment is to increase the
atomic beam flux $R_{p}$, to make the average number of atoms in the
laser mode to be $\overline{N}_{transit}\gg1$[20]. The second
adjustment is to increase the laser cavity length to $4cm$, and
increase the cavity mode waist to $800\mu$m. Then the atom-cavity
coupling constant $g$ is decreased, which is[7],
$g=(\mu/\hbar)\sqrt{2\pi\hbar\omega_{atom}/{V_{mode}}}=2\pi\times1.0kHz$,
where $\mu$ the electric dipole moment, $\omega_{atom}$  the
transition frequency, and $V_{mode}$ the mode volume. The most
probable velocity of atoms in beam $\upsilon_{probable}$ is $505m/s$
while the $^{88}$Sr oven is operating at the temperature of
$630^{\circ}C$. The atomic transit-time through the cavity mode
after averaging over transverse Gaussian profiles[7] is
$t_{transit}=\sqrt{\pi}W_{0}/v=2.8\mu$s . Then the transit-time
broadening[7] is $\Gamma_{transit}/2\pi=4/2\pi t_{transit}=220kHz$.
We have $a\equiv\Gamma_{cavity}/\Gamma_{gain}=50$ while the cavity
decay rate is $\Gamma_{cavity}/2\pi=11MHz$. For
$gt_{transit}=0.018\ll\pi$, the atomic transition probability is
$\sin^{2}(\sqrt{n+1}gt_{transit})$, then the photon emission rate,
i.e., the laser emission coefficient[20] is
$K=\Gamma_{transit}\sin^{2}(\sqrt{n+1}gt_{transit})$ , where $n$ is
the number of photons in laser mode. The rates are shown in Fig.3,
and all parameters of one-atom laser with $^{138}$Ba atom[6,7],
active optical frequency standards with thermal $^{88}$Sr and cold
$^{40}$Ca atoms are listed in Table 1. It will be showed at last
$n\gg1$ , thus the photon-number rate equation will be approximated
by[7,20],

\begin{equation}\label{}
    \frac{dn}{dt}=R_{p}\sin^{2}\left(\sqrt{n}gt_{transit}\right)-n\Gamma_{cavity}.
\end{equation}

The  steady-state solution of Eq.(2) can be written in a
dimensionless form[20],

\begin{equation}\label{}
    r_{\eta}=\frac{n_{\nu}}{\sin^{2}\sqrt{n_{\nu}}},
\end{equation}

with

\begin{eqnarray}
  n_{\nu} &\equiv& n\left(gt_{transit}\right)^{2}, \nonumber\\
  r_{\eta} &\equiv& \frac{N_{transit}}{N_{threshold}}, \nonumber\\
  N_{threshold} &\equiv& \frac{\Gamma_{cavity}}{g^{2}t_{transit}},\nonumber\\
  N_{transit} &\equiv& R_{p}t_{transit},
\end{eqnarray}

Where $N_{transit}$  is the number of atoms in the cavity mode, and
$N_{threshold}$ is the threshold atom number for lasing, thus
$r_{\eta}$ has the meaning of the pumping parameter of conventional
laser. In order to decrease the quantum-limit laser linewidth, set
$r_{\eta}=2$. Around $r_{\eta}=2$, the solution of Eq.(3) is shown
in Fig.4.

\begin{figure}
  \includegraphics[width=7cm]{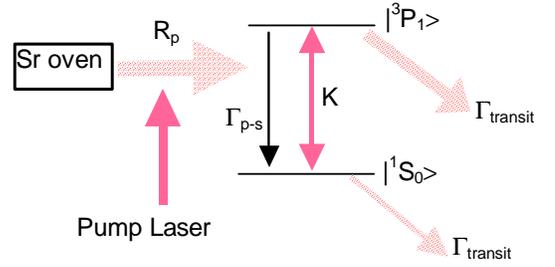}
  \caption{Rates of atomic beam gas laser. Laser pumped atoms
are injected into laser cavity at rate of Rp, the excited state
decay rate $\Gamma_{p-s}$, the transit-time broadening
$\Gamma_{transit}$, and the laser emission coefficient is
$K$.}\label{fig:3}
\end{figure}

From Eq.(4), an atomic flux of $R_{p}=4.3\times10^{11 }atoms/s$ is
needed to satisfy $r_{\eta}=2$ . With the solution of
$n_{\nu}\approx2$ at $r_{\eta}=2$ showed in Fig.4, from Eq.(4), the
steady-state number of photons in laser cavity is
$\overline{n}_{cavity}=n_{\nu}/(gt_{transit})^{2}=6.2\times10^{3}$,
hence the output power of laser $P=\overline{n}_{cavity}h \nu
\Gamma_{cavity}$ is $0.12\mu W$.

The laser linewidth of Eq.(1), at zero detuning will be reduced to,

\begin{eqnarray}
    \Delta\nu_{laser}=\frac{\Gamma_{cavity}}{4\pi\overline{n}_{cavity}}\frac{\left(1+r_{\eta}\right)}{2}\left(\frac{1}{1+a}\right)^{2}.
\end{eqnarray}

Putting all the $^{88}$Sr numerical parameters into Eq.(5), we have
$\Delta \nu_{laser}=0.51Hz$.

As a laser light source, this 0.51Hz linewidth almost reaches the
best-known 0.16Hz linewidth of a cavity stabilized laser[11], which
is achieved within a  $9m^{3}$ wooden enclosure lined internally
with lead foam[21]. As an optical frequency standard, this 0.51Hz
linewidth surpasses the recorded narrowest 6.7 Hz linewidth of Hg
ion clock[14] and measured 27Hz linewidth of most recent optical
lattice clock[5].

When  $N_{sp}$ is expressed in $r_{\eta}$, the Eq.(5) agrees with
the result $\Delta \nu_{laser}=\Gamma_{cavity} (1+\theta^{2})/(8\pi
\overline{n}_{cavity})$ with $\theta^{2}=r_{\eta}$ from quantum
theory of micromaser[22]where the bad cavity effect is not included.
The text-book Schawlow-Townes formula[16,17] is valid only within
the good-cavity regime, which gives a standard text-book example: a
gas laser with milliHertz linewidth. Unfortunately, due to the
vibrations of cavity length, this text-book example of milliHertz
linewidth[16,17] has never been achieved. At the bad-cavity regime,
as the first and second quadrants shown in Fig.1, the linewidth of a
laser or maser will be further modified to be much smaller than the
good-cavity Schawlow-Townes limit with a factor of $(1+a)^{-2}$  as
shown in Eq.(1), and results in the original Schawlow-Townes
formula[1,8]. The elegant experiments of this bad-cavity effect on
laser linewidth have been performed thoroughly in small gas laser in
Woerdman group recently, with clear theoretical explanation[8].

\begin{figure}
  \includegraphics[width=7cm]{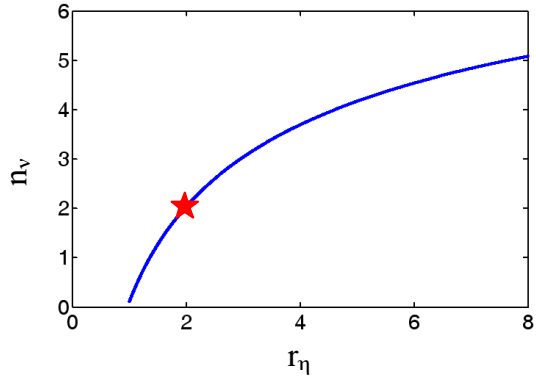}
  \caption{Normalized form of the semi-classical rate equation
solution. Here $r_{\eta}$ has the meaning of the pumping parameter
of conventional laser. The solutions around $r_{\eta}=2$ provide
narrowest laser linewidth.}\label{fig:4}
\end{figure}

The center frequency of a good-cavity gas laser or a super-cavity
stabilized laser follows the cavity length variation almost
perfectly to the level of mHz, and even more tightly is
possible[10-12]. Thus the final technical limitations on the
available laser linewidth are from the variations of the cavity
length[10-12] caused by the environmental vibrations, thermal
expansion, body-force which produces distortion, long-term creep,
and thermal Brownian motion noise[23].  This formidable hurdle is
cleared here by the bad-cavity effect. In the bad-cavity regime, the
laser center frequency doesn't follow the cavity length variation
exactly[8], but in a form of "cavity pulling" shift, which is a
well-known shift in Hydrogen maser[2,24],
$\Delta\nu_{cavity-pulling}=(\Gamma_{gain}/\Gamma_{cavity})\left(\nu-\nu_{0}\right)$.
It is $\Delta \nu_{cavity-pulling}=2\times10^{-2}(\nu-\nu_{0})$ with
the $^{88}$Sr atomic beam value of $a=50$. When the cavity spacer
consists of Zerodur or ULE(Ultra-Low Expansion) and optically
contacting to cavity mirrors, $\pm0.2Hz/s$ cavity mode
drift[3,10-12] only causes a $\pm4mHz/s$ shift of the laser
center-frequency.

Technically, we have set $r_{\eta}=2$ before.  By Eq.(3) and Fig.4,
one can get a larger photon number $\overline{n}_{cavity}$ by
increasing the pumping parameter $r_{\eta}$, but the disadvantage is
spontaneous-emission factor will go up too. It is a technical
trade-off between  $\overline{n}_{cavity}$ and $N_{sp}$ to minimize
the laser linewidth in a practical set-up. Since an $^{88}$Sr atomic
beam of $R_{p}=2\times10^{12}atoms/s$ flux has been achieved[25] at
the oven temperature of $630^{\circ}C$, it is possible to reach much
higher flux while 2cm nozzle array to satisfy the $r_{\eta}=2$
requirement on high atomic flux. When the inhomogeneous broadening
is close to the homogeneous broadening[26], the laser linewidth of
Eq.(1) will increase by a factor of 3. The relative motion between
atoms and cavity in the direction of cavity mode axis due to
vibrations can be neglected. The transverse velocity distribution of
thermal atomic beam will cause the inhomogeneous broadening of gain
profile. It is predicted two-photon Doppler cooling of $^{88}$Sr via
$(5s)^{2}$ $^{1}S_{0}$-5s6s $^{1}S_{0}$ transition can achieve a
Doppler limit of $57\mu K$[27], this means a transverse velocity of
$v_{tranverse}=0.075m/s$, and a narrowed inhomogeneous broadening of
$2\pi\times108kHz$, which is less than the transit time broadening
$\Gamma_{transit}=2\pi\times200kHz$ of laser cavity. Given there is
0.2 micro-radian angle deviation between the 1D transverse cooling
beam and the laser cavity axis, for atomic beam with velocity of
$505m/s$, will result in a line broadening of
$\Delta\nu_{1st-Doppler}=146kHz$.

\begin{table}
\caption{\label{tab:table1}Parameters of one-atom laser and active
optical frequency standards.}
\begin{ruledtabular}
\begin{tabular}{lrrr}
  Laser&One-atom\footnote{The parameters of one-atom $^{138}Ba$ laser are averaged over
  the standing wave transverse Gaussian proofile\cite{7}. But for the parameters of
  $^{88}Sr$ and $^{40}Ca$ active optical frequency standards only the transit times
  and the transit-time broadening are averaged over the transverse
  Gaussian profiles.}&Atomic-beam&Cold-atom\\
  \hline
  Gain medium & $^{138}Ba$ & $^{88}Sr$ & $^{40}Ca$ \\
  $\Gamma_{p-s}/2\pi(kHz)$ & 50 & 7.6 & 0.32 \\
  $g/2 \pi(kHz)$ & 300 & 1.0 & 0.205 \\
  $\upsilon_{probable}(m/s)$ & 360 & 505 & 10 \\
  $t_{transit}(\mu s)$ & 0.2 & 2.8 & 143 \\
  $\Gamma_{transit}/2 \pi (kHz)$ & 3,100 & 220 & 4.3 \\
  $\Gamma_{cavity}/2 \pi(MHz)$& 0.15 & 11 & 0.22 \\
  $a\equiv\Gamma_{cavity}/\Gamma_{gain}$ & 0.05 & 50 & 50 \\
  $K=g^{2}t_{transit}(s^{-1})$ & $3.2\times10^{5}$& 110 & 237 \\
  $R_{p}$(atom/s) & $2.3\times10^{7}$ & $4.3\times10^{11}$ & $8.2\times10^{7}$ \\
  $2^{nd}$ Doppler(Hz)& 270 & 615 & 0.25 \\
  Photons in cavity & 15 \footnote{In the recent $^{138}Ba$ experiment\cite{20},
  the number of photons in the cavity has reached more than 2500, it means
  the laser quantum-limit linewidth is 17Hz at good cavity regime.}& 6,200 & 60 \\
  $P_{output-power}$(nW) & 0.004 & 120 & 0.023 \\
  $\Delta\nu_{laser}(r_{\eta}=2)(Hz)$ & 7,500 & 0.51 & 1.1 \\
\end{tabular}
\end{ruledtabular}
\end{table}

\begin{table}
\caption{\label{tab:table2}Estimated major corrections and
uncertainties of the thermal $^{88}Sr$ beam active optical frequency
standard. All values are in Hz.}
\begin{ruledtabular}
\begin{tabular}{lrr}
Effect &Correction&Uncertainty\\
\hline
  $2^{nd}$-order Doppler & 615& 0.25 \\
  Light shift & 8& 0.08 \\
  Recoil shift& $4,737$ & $1\times10^{-3}$ \\
  $1^{st}$-order Zeeman & 0 & 0.02 \\
  $2^{nd}$-order Zeeman & 4 & 0.04 \\
  Blackbody shift & 1& 0.01 \\
  Collision shift & 0.08 & 0.01 \\
  cavity pulling & 0.1& 0.01 \\
  Recoil asymmetry& 0.02 & $1\times10^{-3}$ \\
  Total  uncertainty &  & 0.27 \\
\end{tabular}
\end{ruledtabular}
\end{table}

An accuracy of this thermal $^{88}$Sr active optical clock is
estimated as follow, and the estimated major corrections and
uncertainties are listed in Table II. The atom density in laser
cavity is $6\times10^{7}atoms/cm^{3}$ in the thermal $^{88}$Sr beam
case, gives 0.08Hz density-dependent frequency shift with the
measured cold $^{88}$Sr atoms result[19]. The recoil-induced shift
of stimulated emission $-4.7kHz$ can be corrected with an
uncertainty less than $1mHz$[19]. For the standing wave in the laser
cavity, the residual first-order Doppler effect only broaden the
line symmetrically, it does not cause shift of laser frequency[28].
The second-order Doppler broadening of $1.6kHz$ can be neglected,
but its asymmetry contributes to frequency uncertainty of clock.
Assuming there is $0.1^{\circ}C$  uncertainty of oven temperature,
the uncertainty of induced second-order Doppler broadening 0.25Hz is
set as frequency uncertainty. The light shift induced by stray light
from pumping laser is estimated to be $8Hz$ providing one tenth of
the pumping laser intensity ($\pi$-pulse for moving atoms) goes into
the cavity and the effective frequency detuning of the stray light
is $1Hz$ supposing the pumping laser is locked to the output laser
of active optical clock finally. Its uncertainty is set to be one
percent of this light shift. The recoil effect will cause the
asymmetry of lineshape on the broad gain background. Considering the
$\pm4.7kHz$ recoil frequency positions and the $220kHz$ gain
bandwidth, we set a frequency correction of $0.02Hz$. When the
cavity mode detuning is $5Hz$, the cavity pulling is $0.1Hz$, and
the cavity pulling may be calibrated to an uncertainty less than
$0.01Hz$. For $0.1^{\circ}C$ oven temperature fluctuation during one
second sampling-time, the corresponding instability of clock is
$100mHz$ from the main source due to the Second-Doppler effect,
which gives a limited Allan variance of
$\sigma(\tau)=2.3\times10^{-16}/\sqrt{\tau}$.

The parameters of cold $^{40}$Ca atom listed in Table I show the
second-order Doppler shift decreases dramatically, and the 0.4ms
long lifetime of $^{3}P_{1}$ state of $^{40}$Ca atom allowing the
pumping laser to be put far away from the cavity to minimize the
light shift induced by the stray light of pumping laser beam. Thus
one can expect an absolute frequency uncertainty will be less than
10mH with cold atoms. Particularly, once atoms trapped in Lamb-Dick
regime of optical lattice with ``magic wavelength'' trapping
laser[5], the limits from Doppler effect on instability and
uncertainty may be almost eliminated, in this case, the active
optical frequency standard based on the lattice atoms can be called
optical lattice laser[29].

Another conceivable scheme is the ``two-photon active optical
clock'', the most attractive one will be ``Hydrogen 1S-2S two-photon
active optical clock'' combining the Hydrogen 1S-2S two-photon
spectroscopy[30] with the two-photon laser[31] under the principles
and techniques presented in this letter.

This active optical clock can use any ``free'' medium: neutral
atoms, ions and molecules. It's expected the extension of the
principles and techniques of active optical clock presented in this
letter will have a great effect on fundamental physics such as
Lorentz invariance test and gravitational wave detection, not only
limited to precision laser spectroscopy and optical atomic clocks.

The author thanks Yiqiu Wang, Donghai Yang, Kaikai Huang, and Xuzong
Chen for helpful discussions. Discussions with Mark Notcutt, Jun Ye
and John L. Hall on ultra-stable cavity are gratefully acknowledged.
This work is founded by MOST under Grand No.
2005CB724500 and NSFC under Grand No. 60178016.\\

\end{document}